\documentstyle[twocolumn,12pt,a4]{article}

\begin{document}
\voffset -3cm
\textheight 25cm
\renewcommand{\thesection}{\arabic{section}.}
\begin{titlepage}
\begin{quote}
\raggedleft  {\bf LMU-91/03} \\ {\em August 1991}
\end{quote}
\vspace{2cm}
\begin{center}
{\bf {GAUGE CREATED TOP QUARK CONDENSATE \\ AND HEAVY TOP}}
\footnote{supported by Deutsche Forschungsgemeinschaft project Fri 412/12-1}
\end{center}
\vspace{2cm}
\begin{center}{\bf {Ralf B\"ONISCH}} \\
\vspace{1cm}
Ludwig-Maximilians-Universit\"at M\"unchen, Sektion Physik  \\
8000 M\"unchen 2, F.R.G.
\end{center}
\vspace{2cm}
\begin{abstract}
The scalar interaction term of the top mode standard model
can be generated by a
gauge theory. It is emphasized that in this approach
generalization to family space directly
yields a heavy top quark while leaving all other quarks massless.

By generating the new interaction from gauge extension to custodial $SU(2)$,
the standard model bound $m_{top} \stackrel{<}{\sim} 200 GeV$ is pushed to
$m_{top} \stackrel{<}{\sim} 270 GeV$, thereby significantly reducing the
amount of finetuning required to match this number.
\end{abstract}
\vspace{2cm}
\end{titlepage}
\renewcommand{\thefootnote}{\fnsymbol{footnote}}

The 1990's experiments have provided us not only with a confirmation of the
fermion-boson sector of the standard model (SM) in a new range of precision,
but also with raised lower bounds on the two missing particles, the top quark
and the Higgs boson and, with the same amount of significance, a surviving
lack of evidence of the scalar sector.

The lower bound of the top mass, $89 GeV$, implies a strong Higgs force,
$g_{t} \simeq {\cal O}(1)$,
and may cause a new possibility of symmetry breaking
in the context of minimal SM: multiple Higgs exchange could bind two tops
to a scalar $\bar{t}t$, which in turn serves as an equivalent of the
elementary massive Higgs boson. Symmetry breakdown might then be driven by
the $\bar{t}t$ condensate, while the Higgs Potential stays non-degenerate
\cite{claro}.

However, the SM scalar sector possesses uncomfortable features: elementary
scalar fields and a suitcase full of free parameters. Therefore any invitation
to replace this part of the theory by something more ``physical" or more
``dynamical" is welcome.
The hope of being able to explain the pattern of fermion masses in detail,
especially the observed hierarchies, is far going as long as there is no
information available on the substructure of these particles.
Nevertheless one can try to make progress by postulating new physics behind
the Higgs sector.

Initiated by the large coupling of Higgs to top, a number of authors have
studied a top quark condensate in the spirit of the Nambu-Jona-Lasinio (NJL)
model [2-11].
The simplest version of this model works with an interaction of the type
$g\bar{\psi}_L\psi_R\bar{\psi}_R\psi_L$.
The particle described by $\psi$ is massless at the tree-level and achieves
a mass via the Schwinger-Dyson equation
\begin{equation}
m = 2 g m \frac{i}{2\pi^{4}} \int {\frac {d^{4}p}{p^{2}-m^{2}}}.
\end{equation}
A non-trivial solution of this equation requires g to exceed a certain
critical value.
Here it is essential to {\em assume} that the mass does not vanish. Eq. (1)
is called the self-consistency equation (or, in analogy to BCS theory, the gap-
equation).
As a direct consequence of such a solution of eq. (1), a massive scalar
bound state $\bar{\psi}\psi$ as well as a massless pseudoscalar bound state
$\bar{\psi}\gamma_{5}\psi$ is formed. The physics of these bound states may
serve as a symmetry breaking sector. In order to break the electroweak
symmetry,
two charged Goldstone bosons are necessary in addition and in connection with
the obvious special role of the top quark, the SM Higgs sector may be replaced
by the NJL interaction of left-handed doublets and right handed tops:
\begin{equation}
\bar{\Psi}_{L} t_{R} \bar{t}_{R} \Psi_{L},
\end{equation}
where $\Psi = (t,b)^{T}$, \cite{bhl}.

This model was recently entitled the top mode standard model (TMSM).
All masses other than the top are there generated just as in the original SM,
i.e. by interaction with the massive $\bar{t}t$ (playing the role of the Higgs)
and a set of appropriate parameters.
The relation of the NJL model to the Yukawa theory is discussed for example in
\cite{lur}, while Hasenfratz et al. \cite{hase}
and Zinn-Justin \cite{zj}
recently argued that the top mode SM is not
a restricted version of the SM but only another parametrization.

However, the scalar
4-fermion interaction (2) will in any case not be the last word.
The physics that might be recovered behind the NJL model should also give some
insight into the mass spectrum of the fermions. In gauge theories chiral
symmetry arguments suggest parity conserving vector boson exchange to produce
fermionic masses, as was discussed by Fritzsch \cite{f}.

So motivated, it was speculated by the author that an additional gauge symmetry
is responsible for the bootstrap physics
\cite{boe}.
On the effective low energy level,
we have an equivalence of the massive scalar exchange and part of the
exchange of a massive vector boson, expressed in the identity
\begin{equation}
\bar{\psi}_{1L}\gamma^\mu\psi_{2L}\bar{\psi}_{3R}\gamma_\mu\psi_{4R}=
-2\bar{\psi}_{1L}\psi_{4R}\bar{\psi}_{3R}\psi_{2L}.
\end{equation}
Thus, if $\psi's$ are multiplets, eq. (3) carries some group structure
appearing
as generators in the couplings. Extended electroweak gauge groups produce
specific hierarchies in the weak doublets as a consequence of the structure of
the additional factor and its mixing behaviour with the $SU(2)_{L} \times
U(1)_{Y}$
gauge fields.

This vertical hierarchy was studied in ref. \cite{boe}.
The mass splittings in the
third family were found to be directly reproduced in the effective couplings of
the custodial $SU(2)$ extension of the SM.\footnote
{For a detailed discussion of this model, see refs. \cite{ks,bmw}.}
The
purpose of this letter is to
update the current discussion on this bootstrap scheme.
For this, we complete the hierarchy in the horizontal direction
by recalling the formalism of flavour democracy
\cite{fpl}
and emphasizing its natural evidence in the present class of models.
Secondly, we want to show that upper
limits on the top mass, obtained by indirect
determination via radiative corrections are relaxed from the SM value of
$\sim 200 GeV$ to significantly higher $\sim 270 GeV$ in the specificly
interesting case of extension to the custodial $SU(2)$.
This is particularly important, because the bootstrap mechanism favours a
rather
large dynamical mass, which collides with the indirect data and can so far only
be accomodated by an otherwise unmotivated amount of finetuning the parameters
in eq. (1).
\vspace{1cm}
\newline
{\em Flavour democracy in the gauge generated bootstrap picture.}
Let us trace the coupling in the NJL interaction term back to the gauge vertex
from which it will be generated via eq. (3). The lagrangian for 3 flavours is
given by
\begin{equation}
{\cal L}_{G}= -g \sum_{i=1}^{3} J_{i}^{\mu} G_{\mu}
            = -g \sum_{i=1}^{3} \bar{\psi}_{i} \gamma^{\mu} \psi_{i} G_{\mu}.
\end{equation}
Our consideration does not depend on the gauge group, so that $G_{\mu}$ is any
kind of gauge field. We take $G$ to be
a neutral field first and come back to charged
fields and currents later. The coupling g depends on the quantum
numbers  of $\psi$, but does not depend on the generation index $i$.
${\cal L}_{G}$ is the lagrangian for three flavours like, say, up, charm
and top.
The low energy efffective current-current interaction is
\begin{equation}
  \sum_{i,j=1}^{3} -\frac{1}{2m_{G}^{2}}
                   \bar{\psi}_{i} \gamma^{\mu} \psi_{i}
                   \bar{\psi}_{j} \gamma_{\mu} \psi_{j}.
\end{equation}
With the aid of eq. (3) the left-to-right part of eq. (5) becomes
\begin{equation}
  \sum_{i,j=1}^{3} \frac{1}{m_{G}^{2}}
                   (\bar{\psi}_{iL} \psi_{jR})
                   (\bar{\psi}_{jR} \psi_{iL}).
\end{equation}
One of the pairs in brackets is the mass term, running on the line of the
diagram, which enters the self-consistency equation, the other
forms the condensate.
Because g does not depend on $i$, equal masses for up, charm and top are
introduced, when
$i,j=1,2,3$. The essential hypothesis of the bootstrap mechanism
was emphasized before to be a nonvanishing $m$
in eq. (1) to be cancelled from both sides. In order to suppress masses of the
first and second generations, we could assume that this is only true
for the top, but we have no reason for that. Instead, it is
straightforward to assume that all possible mass terms are generated, since
massless flavours are identical anyhow and discrimination is purely artificial.
This means eq. (1) has non-trivial solutions, when
identical particles are in the brackets, like in eq. (6).
For the $q=2/3$ quarks there are the proper mass terms $\bar{u}u$, $\bar{c}c$
and $\bar{t}t$, as well as the mixing terms $\bar{u}c$, $\bar{u}t$ etc., giving
just a democratic mass matrix
\begin{equation}
M_{d}=m D, \hspace{1.5cm}
D  = \left( \begin{array}{ccc}
                          1 & 1 & 1   \\
                          1 & 1 & 1   \\
                          1 & 1 & 1    \end{array} \right)
\end{equation}
which diagonalizes to
\begin{equation}
M_{h}=m H, \hspace{1.5cm}
H  = \left( \begin{array}{ccc} 0 & 0 & 0   \\
                               0 & 0 & 0   \\
                               0 & 0 & 1    \end{array} \right)
\end{equation}
by a unitary transformation $U$
\begin{equation}
H=U\frac{1}{3}DU^{-1},
\end{equation}
\vspace{1cm}
\begin{equation}
U  = \left( \begin{array}{ccc}
\frac{1}{\sqrt{2}} & -\frac{1}{\sqrt{2}}  & 0   \\
\frac{1}{\sqrt{6}} & \frac{1}{\sqrt{6}} & -\frac{2}{\sqrt{6}}     \\
\frac{1}{\sqrt{3}} & \frac{1}{\sqrt{3}} & \frac{1}{\sqrt{3}}
  \end{array} \right).
\end{equation}
Thus the theory possesses one massive fermion, to be identified as the top,
while the others are massless.
The mass matrix (7) is the starting assumption of the work of Kaus and Meshkov
\cite{kme,kmes} and appears here as a simple consequence of degeneracy in
the families: both bare masses (assumed to be zero here) and scalar couplings
do not single out a direction in family space.

A charged gauge boson leads to the same situation, although the way mixing
terms are introduced is different:
the effective
interaction of currents with charges $\alpha$ and $\beta$
$  \sum_{i,j=1}^{3} -\frac{1}{m_{G}^{2}}
                   \bar{\psi}_{i}^{\alpha} \gamma^{\mu} \psi_{i}^{\beta}
                   \bar{\psi}_{j}^{\beta} \gamma_{\mu} \psi_{j}^{\alpha}$,
transforms into the ``feeding" term
$  \sum_{i,j=1}^{3} \frac{2}{m_{G}^{2}}
                   \bar{\psi}_{iL}^{\alpha} \psi_{jR}^{\alpha}
                   \bar{\psi}_{jR}^{\beta} \psi_{iL}^{\beta}$
and the arguments from above still hold.
\vspace{1cm}
\newline
We note that our generalization of the bootstrap hypothesis excludes coloured
bound states, like they could have been introduced as a consequence of the
colour singlet nature of electroweak quark currents and the rearrangement,
which gives for colour indices $a,b$ a scalar term
\begin{equation}
\bar{\Psi}_{L}^{a} t_{Rb} \bar{t}_{R}^{b} \Psi_{La},
\end{equation}
The bootstrap mechanism thus works independently in every colour channel.
The above idea of gauge generated bootstrap has very recently been considered
with the ambition to restore the colour structure of the model of ref.
\cite{bhl},
which can be introduced
by an extraordinary right-handed top and left-handed top-bottom
doublet in a less homogeneous gauge structure than considered so far
\cite{hi}.
\vspace{1cm}
\newline
{\em Custodial $SU(2)$ and top mass limit.}
Let us now turn to the specific gauge group
$SU(2)_{L} \times U(1)_{Y} \times SU(2)_{V}$. It reproduces the (vertical)
mass splitting in the weak doublet without any further assumption
\cite{boe}.
Coming from the SM, we are interested in the influence of the new physics on
the
$m_{top}$-bound, which is obtained by indirect determination via radiative
corrections \cite{smrad}.
In the SM we have an upper bound of $m_{top} \stackrel{<}{\sim} 200 GeV$ and it
requires a very large amount of finetuning in the self-consistency equation (1)
to match this number: the scale of new physics is a very high scale of
${\cal O}(10^{15} GeV)$ and implies a physical desert over more than ten orders
of magnitude \cite {bhl}.
This is so far a general feature of the TMSM, although explicit numbers depend
on the methods used in calculations and their reliability.
For this problem to be avoided one expects the $m_{top}$-limit to be relaxed
by the symmetry breaking mechanism itself.
Explicitly, in the present case of a new hidden interaction,
there should in the full theory remain a deviation from the standard
electroweak physics that allows for a higher top mass even when the ratio
of the Fermi scale to the new scale approaches zero.
This non-decoupling is indeed one of the special properties of the
$SU(2)_{V}$-part within the
$SU(2)_{L} \times U(1)_{Y} \times SU(2)_{V}$ gauge theory and was recently
subjected to a numerical study by Kneur, Kuroda and Schildknecht
\cite{kks}.

In order to clarify the relation of the general
$SU(2)_{L} \times U(1)_{Y} \times SU(2)_{V}$ model considered there to the
presently interesting kind of model, we recapitulate the modification of the SM
by discriminating four steps:
\begin{description}
\item [A.] Standard model.
\item [B.] Top mode standard model. The Higgs sector of the SM is replaced
by the effective interaction and the bootstrap hypothesis like described above.
As there is a one-to-one mapping between the models
\cite{hase,zj}, predictions, including indirect $m_{top}$ determination, are
identical in both of them.
\item [C.] Extended gauge group $SU(2)_{L} \times U(1)_{Y} \times SU(2)_{V}$.
We take the mass of the new bosons to be large compared to the electroweak
scale, $m_{V} \gg G_{F}$.
At $G_{F}$, the $V$-exchange is pointlike and the full theory at electroweak
energies can be splitted into the SM plus effective low energy $V$-physics:
\begin{equation}
{\cal L}={\cal L}_{SM}^{\prime} + {\cal L}_{eff}(V).
\end{equation}
Note that ${\cal L}_{SM}^{\prime}$ differs from the pure SM by effects
which do not decouple \cite{ks}.
$V$-loops shall be neglected in radiative corrections of standard processes,
since the coupling is controlled by mixing of the gauge fields
(all fermions have quantum number $T_{3V}=0$ and hence do not couple directly
to the new bosons)
and expected to be small already at the tree level
\cite{bkoe}.
Altogether, radiative corrections are SM radiative corrections to a very
good approximation and deviations from the SM are given by the influence
of the non-decoupling effects that yield the replacement
${\cal L}_{SM} \rightarrow {\cal L}_{SM}^{\prime}$.

This non-decoupling stems from the contribution of gauge coupling of the field
$V_{3}$ to the electric charge via mixing with the standard neutral sector
and results in a modification of the SM relation among the mixing angle
$\theta_{W}$, weak gauge coupling $g$ and electric charge $e$,
$\sin^{2}\theta_{W} g^{2} / e^{2} = 1$, to
\begin{equation}
\sin^{2}\theta_{W} g^{2} / e^{2} + \epsilon = 1,
\end{equation}
where $\epsilon$ parametrizes the non-standard contribution.
Expressing $\epsilon$ in terms of $e,G_{F}$,
\linebreak
$\sin^{2}\theta$ and $m_{Z}$
and, with the arguments from above, inserting SM radiative corrections,
yields the desired relation of the non-standard effects and the top mass, which
enters the loop calculations: $\epsilon=\epsilon(m_{top})$.
This relation was derived in \cite{kks} and we shall quote the
numerical result.
\item [D.] Bootstrap from $\Lambda_{V}$ to $G_{F}$.
We start at step C and make the replacement that lead from step A to step B.
The pure SM part in ${\cal L}_{SM}^{\prime}$, eq. (12), is modified to the
TMSM,
while the 4-fermion term is now generated by effective $V_{0}$-exchange.
Clearly, ${\cal L}_{eff}$ in eq. (12) includes many more 4-fermion terms than
those needed to form the minimal composite SM Higgs sector.
(There are also
$\bar{\psi}_{L(R)}\gamma^\mu\psi_{L(R)}\bar{\psi}_{L(R)}\gamma_\mu\psi_{L(R)}$
terms and left-handed charged channels.)
Nevertheless, ${\cal L}_{eff}$ does not contribute in radiative corrections and
steps A and B gave the same results.
Thus deviations of the model step D from the SM are those of step C from the SM
and the result of \cite{kks} is valid also at this step.
There it is shown that for a small negative value of
$\epsilon \simeq -0.01$ the top mass window is easily opened up to
\begin{equation}
m_{top} \stackrel{<}{\sim} 270 GeV.\footnote
{There is no relation between the scalar top channel coupling $G_{t}$,
which has to be positiv, and $\epsilon$.
The accidental appearance of $\epsilon$ in $G_{t}$ is without consequence:
$$ G_{t} \simeq e^{2} \left( \frac{1}{2(1-m_{v}^{2}/m_{v_{0}}^{2})} \right)
                       [1 + \epsilon + {\cal O}(\epsilon^{2})],$$
the sign of $\epsilon$ does not change the sign of $G_{t}$.}
\end{equation}
\end{description}
With this raised upper bound, eq. (14), the TMSM is much less sensitive to
finetuning and the new scale can be very far below a GUT scale.
\vspace{1cm}
\newline
A few words are in order concerning renormalizability.
It was frequently objected that the effort of the top mode standard model,
the avoidance of SM elementary scalar fields, is lost in the present
class of models, since renormalizability is restored by introducing an
appropriate set of scalar fields to break the new gauge symmetry.
However (apart from the findings in
\cite{hase,zj})
these scalars are not responsible for the SM fermion masses, nor for
the masses of W and Z.
Instead, we seek a dynamical model of breaking the {\em electroweak}
symmetry and generating standard
fermion masses.
Thus the new scalar sector is not a source of many
arbitrary parameters.
Moreover, it should be regarded as an advantage that the theory can be made
renormalizable, in the sense that scalars serve as a working description
(just like they do in the SM), instead of remaining in an effective model.
\vspace{1cm}
\newline
To summarize, the custodial $SU(2)$ yields the gross features of the fermion
spectrum, when made responsible for symmetry breaking and mass generation in
the spirit of the NJL model as it was proposed recently
\cite{boe}\footnote
{A presentation of the whole model is in preparation.}.
Remarkably, it possesses some peculiarities, like
the singlet nature of fermions and non-decoupling of heavy gauge bosons, being
potentially connected with inner reasons of mass generation or
able to help valueing this scenario in general.
More evidently, by only one very simple assumption one has a
heavy top quark, while all other quarks are massless and the
lepton sector behaves analogously.
Towards the observed mass values more detailed effects will
have to disturb this rough mechanism.
\vspace{2.5cm}
\newline
I thank J.L.Kneur for a number of useful comments and
H. Fritzsch, R.K\"ogerler, C.Lucchesi, S.Reinshagen and J.Zinn-Justin
for some discussions.
\vspace{0.3cm}
\newline
After this work was finished, I received another paper
\cite{mali}, which is based on the present idea of the bootstrap origin and
also throws some light on the striking advantages of this new class
of models in comparison with technicolour theories.

\end{document}